\pdfoutput=1

\def\40K{$^{40}$K}
\def\K{$^{39}$K}
\def\Na{$^{23}$Na}
\def\NaK{\Na\K}

\def\ket#1{\mathinner{|{#1}\rangle}}

\documentclass[aps,pra,twocolumn,floatfix,10pt]{revtex4} 
\usepackage{graphicx,amsmath,amssymb,ulem}
\usepackage{textcomp}
\usepackage{hyperref}
\usepackage{color}
\usepackage{epstopdf}
\usepackage{nicefrac}
\usepackage[caption=false]{subfig}
\usepackage{gensymb}
\usepackage{placeins}

\begin{document}

\title{Formation of ultracold weakly bound dimers of bosonic \NaK}

\author{Kai~K.~Voges}
\author{Philipp~Gersema}
\author{Torsten~Hartmann}
\author{Torben A.~Schulze}
\author{Alessandro~Zenesini}
\author{Silke~Ospelkaus}
\email{silke.ospelkaus@iqo.uni-hannover.de}

\affiliation{Institut f\"ur Quantenoptik, Leibniz Universit\"at Hannover, 30167~Hannover, Germany}

\date{\today}

\begin{abstract}
We create weakly bound bosonic \NaK\ molecules in a mixture of ultracold \Na\ and \K. The creation is done in the vicinity of a so far undetected Feshbach resonance at about $196\,\text{G}$ which we identify in this work by atom-loss spectroscopy. We investigate the involved  molecular state by performing destructive radio frequency binding energy measurements. For the constructive molecule creation we use radio frequency pulses with which we assemble up to $6000$ molecules. We analyze the molecule creation efficiency as a function of the radio frequency pulse duration and the atom number ratio between \Na\ and \K. We find an overall optimal efficiency of $6\,\%$ referring to the \K\ atom number. The measured lifetime of the molecules in the bath of trapped atoms is about $0.3\,\textrm{ms}$.
\end{abstract}
\maketitle

\section{Introduction}
Feshbach molecule creation from ultracold atomic gases has led to exciting developments ranging from the observation of the BCS-BEC crossover \cite{Chin1128,PhysRevLett.92.040403} to Efimov physics \cite{CsEfi} and state-to-state chemistry \cite{Shanghai2}. Furthermore, Fesh\-bach molecules constitute an important steppingstone for the creation of deeply bound molecules by means of stimulated Raman adiabatic passage (STIRAP). For heteronuclear molecules, this was first demonstrated for fermionic KRb molecules starting from a heteronuclear K and Rb quantum gas mixture \cite{KRb1}. Heteronuclear ground state molecules are of special interest because of their large electric dipole moment. The anisotropic and tuneable dipole-dipole interaction can be used for the control of ultracold chemical reactions \cite{Ospelkaus2010}, quantum simulation \cite{KRbSpin} and quantum computing \cite{DeMilleQuantComp}.\\\noindent
In recent years different bi-alkali heteronuclear ground state molecules have been produced by association of weakly bound dimers and subsequent STIRAP to the ground state. So far, fermionic KRb \cite{PhysRevLett.97.120402,KRb1}, LiNa \cite{GsDiMo23Na6LiBos2012,GsDiMo23Na6Li2017}, NaK \cite{FBMol23Na40K2012,GsDiMo23Na40K2015} and bosonic RbCs \cite{PhysRevA.85.032506,GsDiMo87Rb133Cs2014Grimm} and NaRb \cite{NaRbFeshbach,NaRb3}  have been created in different experiments. Up to now the bosonic molecule \NaK\ is missing, although it might enable an interesting comparison to its fermionic counterpart. NaK ground state molecules have an intrinsic dipole moment of $2.72\,\text{D}$ and are known to be chemically stable against two-body exchange reactions \cite{Inelastic2010}.\\\noindent
In our experiment we aim for the creation of \NaK\ ground state molecules. Therefore, a detailed investigation of weakly bound dimers is mandatory for an efficient ground state molecule production. Recently, Feshbach resonances and refined molecular potentials for the \Na+\K\ mixture have been reported and a quantum degenerate Bose-Bose mixture has been produced \cite{Hartmann2019,SchulzeBEC2018}. Additionally, to transfer the dimers to the ground state, possible STIRAP pathways have been investigated theoretically \cite{Schulze2013} and experimentally \cite{Voges2019}.\\\noindent
Weakly bound Feshbach molecules have mainly been created using two different approaches. First of all, magnetic field ramps have been used mostly in the vicinity of narrow Feshbach resonances. Second, direct state transfer methods have been implemented mostly in the vicinity of broad Feshbach resonances. In this case, the bound molecular state is directly populated starting from a non-resonant scattering channel using radio frequencies (rf), microwave radiation, magnetic field modulation \cite{RevModPhysFesh} or optical Raman transitions \cite{SpinOrbitFeshMole}.
\\\noindent
Here we report the formation of weakly bound \NaK\ dimers from an ultracold mixture of bosonic \Na\ and
\K\ by means of rf association. In our experiment we make use of an up to now undetected broad Feshbach resonance in the $\ket{f=1, m_f=-1}_\text{Na} + \ket{f=2, m_f=-2}_\text{K}$ scattering channel at approximately $196\,\text{G}$. Here $f$ is the hyperfine quantum number and $m_f$ its projection on the magnetic field axis. We locate the Feshbach resonance by magnetic field dependent atom-loss spectroscopy. We also measure the binding energy of the involved molecular state and characterize the efficiency of the molecule creation process.\\\noindent\newline
In the following we describe our experimental procedure in Sec.\ref{ExpProc}. Characterization measurements of the Feshbach resonance are summarized in Sec.\ref{FeshPos}. Finally, we discuss the formation of weakly bound dimers  and the efficiency of the creation process in Sec.\ref{Creation}.

\section{Experimental procedures}
\label{ExpProc}
For the presented experiments the atomic mixture is prepared following the procedure described in \cite{SchulzeBEC2018,Hartmann2019}. We start with two pre-cooled atomic beams, the one for \Na\ produced by a Zeeman slower and the one for \K\ produced by a two-dimensional magneto-optical trap (MOT). Atoms from both beams are captured in a three-dimensional MOT. Afterwards both species are individually molasses cooled before they are optically pumped to the $F=1$ state and loaded into an optically plugged magnetic quadrupole trap. In the trap \Na\ atoms are cooled by forced microwave evaporation. \K\ atoms are sympathetically cooled in the bath of \Na\ atoms. The cold atomic mixture is loaded from the magnetic trap to a $1064\,\text{nm}$ crossed-beam optical dipole trap (cODT). The cODT intensity is increased while the quadrupole field is switched off and a homogeneous magnetic field of about $150\,\text{G}$ is applied yielding favorable inter- and intra-species scattering lengths. A final optical evaporation step is performed by lowering the intensity in both beam of the cODT. For the experiments both species are cooled to temperatures below $1\,\mu\textrm{K}$. The atom number ratio between the two species is adjusted by the depth of the magnetic trap evaporation before the mixture is loaded to the cODT. During the sympathetic cooling process the phase space density of \K\ atoms increases and hence three-body losses on \K\ occur \cite{Schulze2018} reducing drastically the atom number. We can vary the atom number ratio $N_\text{Na}/N_\text{K}$ in the $\ket{f=1, m_f=-1}_\text{Na} + \ket{f=1, m_f=-1}_\text{K}$ channel in the final cODT between 0.3 and 18.\\\noindent

\section{Feshbach resonance in the $\ket{1,-1}_\text{Na}+\ket{2,-2}_\text{K}$ state}
For a successful creation of weakly bound dimers by rf pulses precise knowledge of the involved molecular state is essential. The Fesh\-bach resonance and therefore the molecular state used here have never been observed before. We first investigate the Feshbach resonance by means of atom-loss spectroscopy; see Sec.\ref{AtomLoss}. For the precise determination of the resonance position we perform destructive binding energy measurements as a function of magnetic field; see Sec.\ref{MolBild}.
\label{FeshPos}
\subsection{Atom-loss spectroscopy}
\label{AtomLoss}
\begin{figure}[h]
	\includegraphics[width=1\columnwidth]{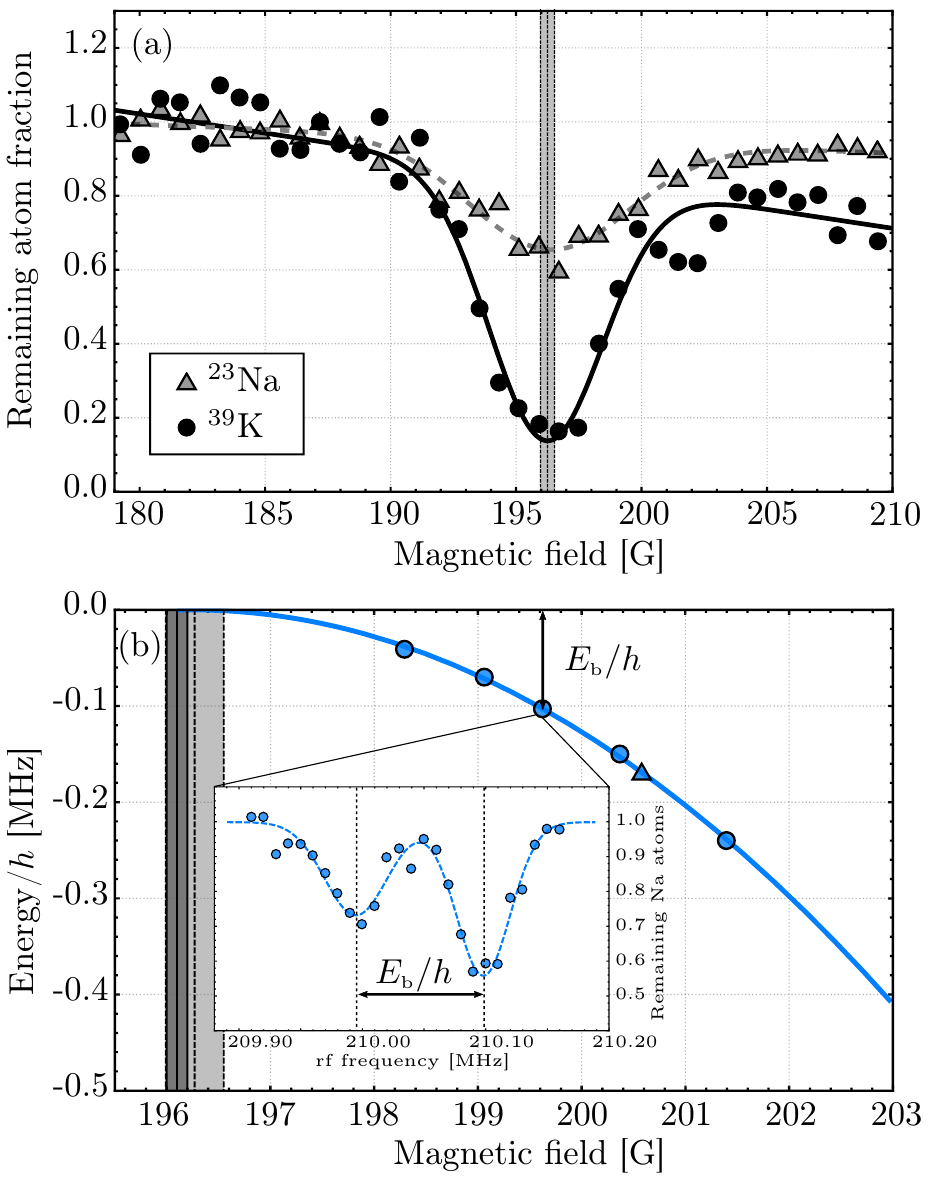}
	\caption{Feshbach resonance characterization. (a) Atom-loss spectroscopy in the $\ket{f=1, m_f=-1}_\text{Na} + \ket{f=2, m_f=-2}_\text{K}$ channel. Remaining atom fraction of \Na\ and \K\ for different magnetic field strengths with a holding time of $40\,\textrm{ms}$. The remaining atom number is normalized to the initial atom number for each species, respectively. Open gray triangles and the dashed gray line refer the to remaining \Na\ atom fraction and full black circles and the solid black line refer to the \K\ atom number fraction. The vertical lines and corresponding shaded area indicates the resonance position and the standard deviation obtained from the two Gaussian fits. (b) Binding energy of the weakly bound state as obtained from rf spectroscopy. The blue continuous line is a fit to the destructive measurements (blue circles) according to universal binding energy Eq.(\ref{EBind}). The triangle is the binding energy extracted from the constructive signal from Fig.\ref{ConSig}. The vertical lines are the resonance position with the standard deviation, dashed for the atom-loss measurement in (a), solid lines for the binding energy measurement.  The inset shows an sample rf scan for a binding energy of $h \times 103\,\text{kHz}$ with a double-Gaussian fit (dashed blue line). The arrows indicate the binding energy. Error bars in both figures are smaller than the plot markers and are not shown.}
	\label{FeshbackData}
\end{figure}
In the vicinity of Feshbach resonances the atoms experience an increased scattering rate which enhances two- and three-body losses. To determine the position of the Feshbach resonance we use this effect and perform atom-loss spectroscopy. The Feshbach resonance of interest  is predicted to be located at a magnetic field of about $196\,\text{G}$ in the $\ket{f=1, m_f=-1}_\text{Na} + \ket{f=2, m_f=-2}_\text{K}$ scattering channel \cite{Hartmannphd}. We prepare the atomic mixture as explained above and finally transfer the \K\ atoms to the $\ket{f=2, m_f=-2}_\text{K}$ state at $137\,\text{G}$ using a rapid adiabatic passage (RAP) \cite{ARPdressedAtomInterpret1984,Hartmann2019}. In this specific case we apply a rf of $256\,\text{MHz}$ and swept the magnetic field by approximately $1\,\text{G}$ within $100\,\textrm{ms}$. At about $137\,\text{G}$ the inter- and intra-species scattering rates for the \Na+\K\ mixture for all involved state combinations are low enough to allow for sufficient long holding time for the RAP \cite{Hartmannphd}.\\\noindent
To start atom-loss spectroscopy of the Feshbach resonance, we perform a fast magnetic field ramp to different magnetic field values in the vicinity of the resonance. We hold the mixture at each magnetic field value for $40\,\text{ms}$ so that the enhanced scattering rate leads to atom losses. We record the remaining atom number after the hold time resulting in a loss feature as shown in Fig.\ref{FeshbackData}(a). Using a phenomenological Gaussian fit to the data of \Na\ and \K\ respectively, we determine the resonance position to be at $196.27(28)\,\text{G}$. The value is in good agreement with predictions obtained from our most recent available NaK ground state potentials \cite{Hartmannphd}.

\subsection{Molecular binding energy}\label{MolBild}
At the position of a Feshbach resonance a molecular state enters from the diatomic continuum into the scattering threshold. As both states are highly coupled close to the Feshbach resonance, the molecular state becomes spectroscopically accessible. To observe the bound state, we perform destructive binding energy measurements starting from a free diatomic state. We start with the atomic mixture in the $\ket{f=1, m_f=-1}_\text{Na} + \ket{f=1, m_f=-1}_\text{K}$ state; see Sec.\ref{ExpProc}, and apply rf radiation to bridge the energy gap to the resonant $\ket{f=1, m_f=-1}_\text{Na} + \ket{f=2, m_f=-2}_\text{K}$ state and the energetically lower lying bound molecular state. The rf radiation in this experiment is switched on for $40\,\text{ms}$. This ensures that atoms, which are transferred to the resonant $\ket{f=2, m_f=-2}_\text{K}$ state and which experience a high number of scattering events, are significantly depleted from the trap. The particles transferred to the weakly bound molecular state also experience a large loss; see inset Fig.\ref{FeshbackData}(b). The obtained data for different magnetic field strengths is shown in Fig.\ref{FeshbackData}(b). The inset is showing an example of a single binding energy measurement at $199.62\,\text{G}$. The atom-loss feature serves as a calibration of the magnetic field. The separation gives access to the molecular binding energy. The smaller depth of the loss signal for the molecular transition can be attributed to a weaker coupling between atoms and molecules than between atoms.\\\noindent
Close to the resonance position and on the positive scattering length side of the Feshbach resonance the binding energy $E_\textrm{b}(B)$ can be described by the universal formula
\begin{equation}
E_\text{b}(B)=\frac{\hbar^2}{2\mu\,a(B)^2}\,, \label{EBind}
\end{equation} with
\begin{equation}
a(B)=a_\text{bg}\left(1-\frac{\Delta}{B-B_0}\right)
\end{equation}
where $a$ is the interspecies scattering length, $a_\text{bg}$ the background scattering length of the entrance channel and $\Delta$ and $B_0$ the width and position of the Feshbach resonance. $\mu$ is the reduced mass of the molecular system. The solid blue line in Fig.\ref{FeshbackData}(b) is a fit according to Eq.(\ref{EBind}). Using a background scattering length of $a_\textrm{bg}=-38.1\,a_0$, $a_0$ being the Bohr radius, as obtained from coupled channel calculation \cite{privcomET}, we determine the resonance width to the zero crossing to be $\Delta=105.8(1.6)\,\textrm{G}$ and the position to be $B_0=196.10(10)\,\text{G}$, being in good agreement with the results from the loss measurement; see Sec.\ref{AtomLoss}.
\FloatBarrier
\section{Weakly bound dimers}
\label{Creation}
\subsection{Creation process}
\label{creation}
\begin{figure}[b]
	\includegraphics[width=1.0
\columnwidth]{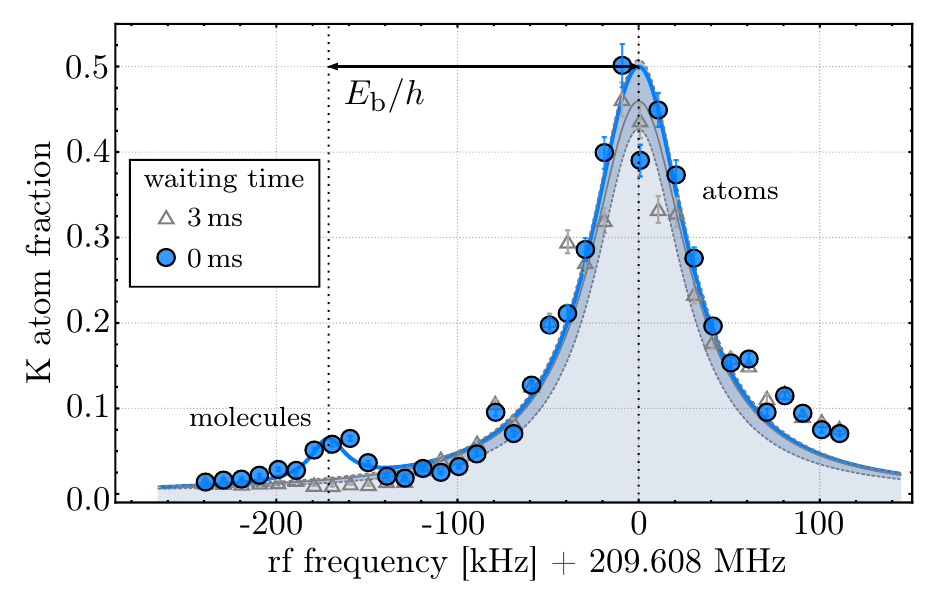}
	\caption{Constructive molecular signal. \K\ atom number as a function the applied rf radiation. The blue circles are the detected atom number directly after the rf pulse. Fitting a phenomenological double-Lorentzian functions (solid blue line) the atomic transition occurs at $209.608(3)\,\text{MHz}$ corresponding to a magnetic field of $200.58(3)\,\text{G}$. The extracted binding energy is $h \times 171(4)\,\text{kHz}$, shown as the black arrow. The value is displayed also in Fig.\ref{FeshbackData}(b) as blue triangle. Gray open triangles are the measured background atoms after an additional waiting time of $3\,\textrm{ms}$ (see text) and the gray solid line with the underlying gray shaded area is a Lorentzian fit with the standard error as darker shaded area. The atom number is normalized to the total number of \K\ atoms. Error bars are the standard deviation and come from different experimental runs.}
	\label{ConSig}
\end{figure}
With the knowledge of the exact resonance position and molecular binding energy, the molecular state can be selectively populated by rf radiation. We start with an atomic mixture in the $\ket{f=1, m_f=-1}_\text{Na} + \ket{f=1, m_f=-1}_\text{K}$ state at a temperature of $500\,\textrm{nK}$ and an atom number ratio of $N_\text{Na}/N_\text{K}=3$. Applying a rf pulse we associate molecules immediately followed by a state selective imaging of the atoms in the $\ket{f=2, m_f=-2}_\text{K}$ state. For this purpose the laser frequency for the \K\ imaging is shifted to be on resonance with the $\ket{S_{1/2}, f=2, m_f=-2}_\text{K} \leftrightarrow \ket{P_{3/2},f=3, m_f=-3}_\text{K}$ transition for a given magnetic field, which is a closed transition from the Zeeman to the Paschen-Back regime and allows to high-field image \K\ atoms at arbitrary magnetic field values. As long as the molecular binding energy is smaller than the linewidth of the atomic transition, this cycling transition can be used to also image weakly bound dimers.\\\noindent
For molecule creation, we switch on the rf source $75\,\textrm{ms}$ before the actual molecule creation takes place with a frequency detuned by $-120\,\textrm{kHz}$ with respect to the molecular transition. We then jump the rf frequency for $0.6\,\textrm{ms}$ to the frequency required for molecule creation followed by a hard  switch-off. We found this method to be more reliable and stable than a simple switch-on/off accounting for less disturbance of the rf to our magnetic field stabilization system. We populate selectively the atomic and the molecular state as shown in Fig.\ref{ConSig}. Both peaks reveal an asymmetric shape on the positive frequency side originating from this pulse application technique; see Fig.\ref{ConSig}. The creation pulse is followed by \K\ imaging as explained above. With this method molecule creation efficiency can be as high as $6\,\%$ at a binding energy of $E_\textrm{b}=h\times100\,\text{kHz}$. For smaller binding energies the atomic and molecular peak start to overlap. In this case it is not possible to prepare pure samples of bound dimers and distinguish them from free atoms in a single experimental cycle. In a separated experimental cycle we introduce a waiting time of $3\,\textrm{ms}$ between molecule creation and imaging to distinguish the short-living molecules from atoms; see triangle symbols in Fig.\ref{ConSig}.

\subsection{Characterization and optimization}
\label{characterization}
\begin{figure}[t]
	\includegraphics[width=1\columnwidth]{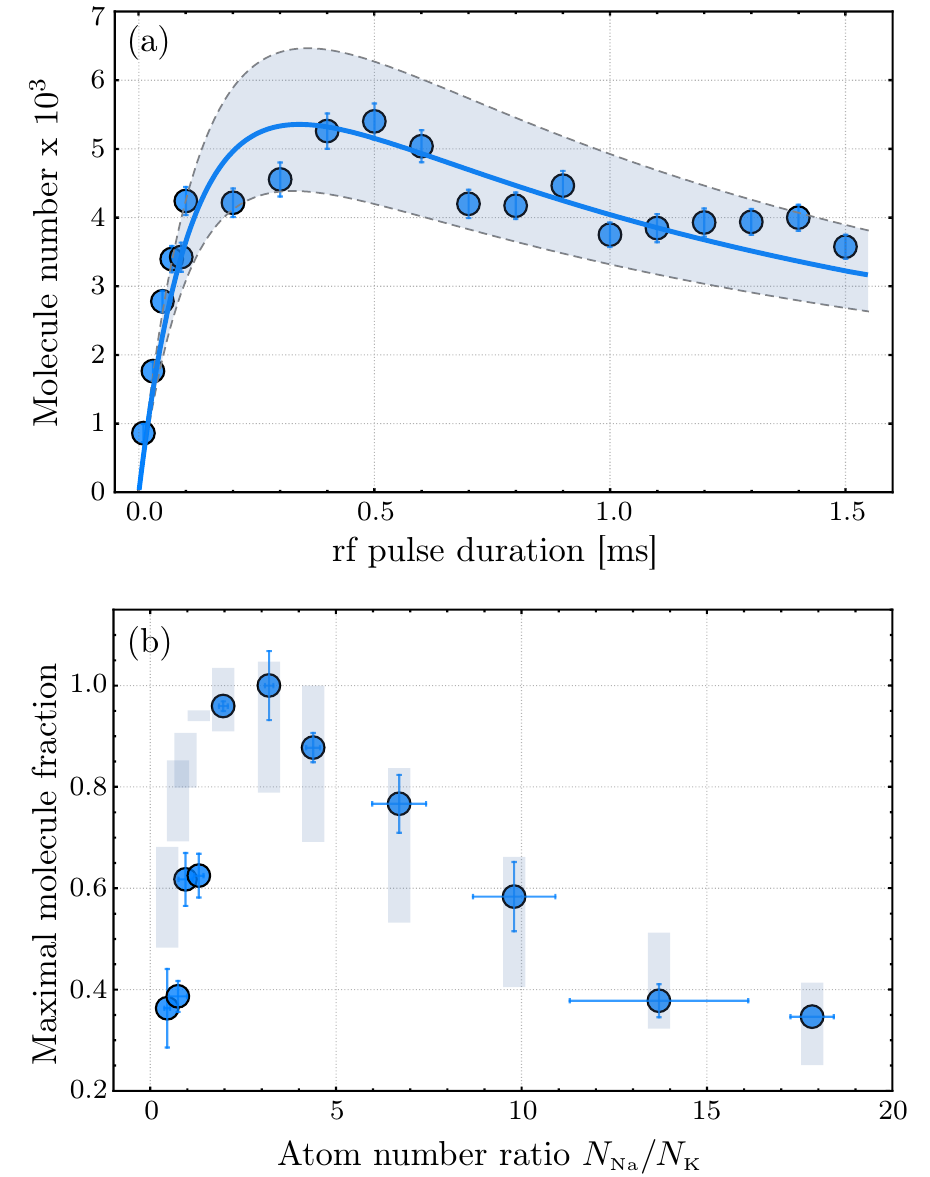}
	\caption{Characterization of the association process. (a) The molecule number is plotted as a function of the rf pulse duration. The molecules are imaged directly after switching off the rf. The best creation efficiency is reached at a pulse duration around $350\,\mu\text{s}$. The solid blue line is the fit modeled with the set of differential equation Eq.(\ref{creationdyn}). The dashed lines with the enclosed shaded area refer to the fit uncertainties. (b) Normalized molecule number as function of the atom number ratio at a creation time of $500\,\mu\text{s}$. The bars represent the predictions from the fit results in (a) taking the individual starting conditions for each point as well as the fit uncertainties into account. The normalization is done according to the maximal created molecule number. Error bars in both figures are the standard deviation and come from different experimental runs.}
	\label{Creationtime}
\end{figure}

For an efficient molecule association the duration of the rf pulse as well as the atom number ratio between \Na\ and \K\ is critical.\\
The dependence on the pulse duration is shown in Fig.\ref{Creationtime}(a). The maximum appears when the molecule creation is overcome by losses of the molecules due to collisions with atoms. The dynamic is modeled by a set of three differential equations for the time-dependent populations $N_\textrm{mol}$, $N_\textrm{Na}$ and $N_\textrm{K}$

\begin{align} \label{creationdyn}
\frac{dN_\textrm{mol}}{dt}=&\,k_\textrm{mol}\cdot g_\textrm{Na,K}N_\textrm{Na}N_\text{K}\nonumber\\&-k_\textrm{a}\cdot (g_\textrm{Na,mol}N_\textrm{Na}+g_\textrm{K,mol}N_\textrm{K})\,N_\textrm{mol}\nonumber\\
\frac{dN_\textrm{Na}}{dt}=&-k_\textrm{mol}\cdot g_\textrm{Na,K} N_\textrm{Na}N_\textrm{K}\\&-k_\textrm{a}\cdot g_\textrm{Na,mol}N_\textrm{Na}N_\textrm{mol}\nonumber\\
\frac{dN_\textrm{K}}{dt}=&-k_\textrm{mol}\cdot g_\textrm{Na,K} N_\textrm{Na}N_\textrm{K}\nonumber\\&-k_\textrm{a}\cdot g_\textrm{K,mol}N_\textrm{K}N_\textrm{mol}\nonumber\,,
\end{align}
where $N_{\textrm{mol}(\textrm{Na})\lbrack\textrm{K}\rbrack}$ are the particle numbers for molecules (\Na\ atoms) $\lbrack\textrm{\,\K\ atoms}\rbrack$, $g_{i,j}$ is the two-body overlap integral \cite{GFesh} and $k_{\textrm{mol}}$ is the molecular creation coefficient and $k_{\textrm{a}}$ the loss coefficient for atom-molecule collisions. The loss coefficient $k_{\textrm{a}}$ is set equal for the case of a colliding molecule either with a \Na\ atom or a \K\ atom. Collisions between molecules are excluded from the model as they are expected to be negligibly small. The solid line in Fig.\ref{Creationtime}(a) shows a fit using the rate model system Eq.(\ref{creationdyn}) with $k_{\textrm{mol}}$ and $k_{\textrm{a}}$ as free parameters. According to the fit, $k_{\textrm{mol}}=1.18(16)\cdot 10^{-9}\frac{\textrm{cm}^3}{\textrm{s}}$ and $k_{\textrm{a}}=4.54(45)\cdot 10^{-9}\frac{\textrm{cm}^3}{\textrm{s}}$. The maximal creation efficiency is found at a pulse duration of about $350\,\mu\textrm{s}$. By accounting the initial atom numbers the loss rate obtained from the second line in Eq.(\ref{creationdyn}) is $k_\textrm{a}\cdot (g_\textrm{Na,mol}N_\textrm{Na}(0)+g_\textrm{K,mol}N_\textrm{K}(0))$ corresponding to a calculated lifetime of $184(23)\,\mu\textrm{s}$.
\\\noindent
We have also measured the molecule formation efficiency as a function of the atom number ratio. We find the highest efficiency at $N_\text{Na}/N_\text{K}\approx3$; see Fig.\ref{Creationtime}(b). We use the parameter from the fit in Fig.\ref{Creationtime}(a) and the specific atom number ratios and total atom number for each experimental data point to calculate the maximal associated molecule fraction. The results are plotted as bars in Fig.\ref{Creationtime}(b) for a direct comparison to the experimental data. Despite the model not accounting for temperature effects such as anti-evaporation or temperature disequilibrium, the predictions are in good agreement with the observed dependence on the atom number ratio.

\subsection{Molecule lifetime}
\label{lifetime}
\begin{figure}[t]
	\includegraphics[width=1\columnwidth]{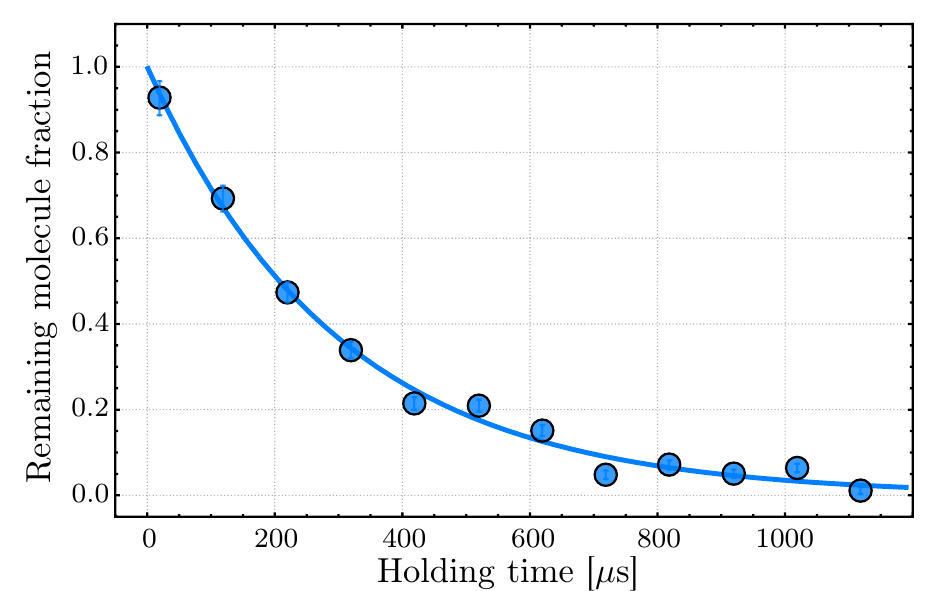}
	\caption{Lifetime measurement of the molecules in the cODT. The molecule number is normalized to the value of zero waiting time. For this measurement \Na\ and \K\ atoms are not removed. A fit to the data (solid line) results in a lifetime of $\tau=299(17)\,\mu\text{s}$. Error bars are the standard deviation and come from different experimental runs.}
	\label{Lifetime}
\end{figure}
We measure the lifetime for molecules immersed in the bath of residual atoms. This measurement is done by introducing a hold time between the rf pulse and the imaging. Figure \ref{Lifetime} shows a typical measurement. For this experiment the creation pulse duration is set to $500\,\mu\text{s}$. During the hold time the rf is set back to the offset detuning to stop molecule creation during the hold time. Using the model from Eq.(\ref{creationdyn}), we determine a lifetime of $299(17)\,\mu\textrm{s}$, which differs from the one obtained in Sec.\ref{characterization}. We suspect that the observed difference of lifetimes originates from the different frequencies of the applied rf during the holding time. This effect is not considered in our model. The observed lifetime is sufficient for subsequent STIRAP transfer to the molecular ground state which is typically 10 to 20$\,\mu\textrm{s}$ long \cite{RevModPhys.89.015006}.
\FloatBarrier
\section{Conclusion and Outlook}
In summary we have investigated a Feshbach resonance for the bosonic \Na+\K\ mixture in the $\ket{f=1, m_f=-1}_\text{Na} + \ket{f=2, m_f=-2}_\text{K}$ channel around $196\,\text{G}$. We located the Feshbach resonance using atom-loss spectroscopy as well as destructive binding energy measurements on the bound molecular state. By applying rf pulses we have been able to populate the bound molecular state and distinguish the dimers from free atoms. We have further characterized and optimized the Feshbach molecule creation efficiency with respect to rf pulse duration and atom number ratio. We have been able to create up to 6000 weakly bound molecules per experimental cycle. The lifetime of the dimers in presence of background atoms is about $0.3\,\textrm{ms}$, which is sufficiently long to perform a STIRAP transfer. These dimers serve as an ideal starting point for efficient creation of so far unobserved ultracold chemically stable bosonic \NaK\ molecules in their absolute electronic and rovibrational ground state.

\section{Acknowledgements}
We thank E. Tiemann for enlightening discussions and theory support. We gratefully acknowledge financial support from the European Research Council through ERC Starting Grant POLAR and from the Deutsche Forschungsgemeinschaft (DFG) through CRC 1227 (DQ-mat), project A03 and FOR 2247, project E5. K.K.V. and P.G. thank the DFG for financial support through Research Training Group 1991.


\begin{thebibliography}{30}%
		\makeatletter
		\providecommand \@ifxundefined [1]{%
			\@ifx{#1\undefined}
		}%
		\providecommand \@ifnum [1]{%
			\ifnum #1\expandafter \@firstoftwo
			\else \expandafter \@secondoftwo
			\fi
		}%
		\providecommand \@ifx [1]{%
			\ifx #1\expandafter \@firstoftwo
			\else \expandafter \@secondoftwo
			\fi
		}%
		\providecommand \natexlab [1]{#1}%
		\providecommand \enquote  [1]{``#1''}%
		\providecommand \bibnamefont  [1]{#1}%
		\providecommand \bibfnamefont [1]{#1}%
		\providecommand \citenamefont [1]{#1}%
		\providecommand \href@noop [0]{\@secondoftwo}%
		\providecommand \href [0]{\begingroup \@sanitize@url \@href}%
		\providecommand \@href[1]{\@@startlink{#1}\@@href}%
		\providecommand \@@href[1]{\endgroup#1\@@endlink}%
		\providecommand \@sanitize@url [0]{\catcode `\\12\catcode `\$12\catcode
			`\&12\catcode `\#12\catcode `\^12\catcode `\_12\catcode `\%12\relax}%
		\providecommand \@@startlink[1]{}%
		\providecommand \@@endlink[0]{}%
		\providecommand \url  [0]{\begingroup\@sanitize@url \@url }%
		\providecommand \@url [1]{\endgroup\@href {#1}{\urlprefix }}%
		\providecommand \urlprefix  [0]{URL }%
		\providecommand \Eprint [0]{\href }%
		\providecommand \doibase [0]{https://doi.org/}%
		\providecommand \selectlanguage [0]{\@gobble}%
		\providecommand \bibinfo  [0]{\@secondoftwo}%
		\providecommand \bibfield  [0]{\@secondoftwo}%
		\providecommand \translation [1]{[#1]}%
		\providecommand \BibitemOpen [0]{}%
		\providecommand \bibitemStop [0]{}%
		\providecommand \bibitemNoStop [0]{.\EOS\space}%
		\providecommand \EOS [0]{\spacefactor3000\relax}%
		\providecommand \BibitemShut  [1]{\csname bibitem#1\endcsname}%
		\let\auto@bib@innerbib\@empty
		\bibitem [{\citenamefont {Chin}\ \emph {et~al.}(2004)\citenamefont {Chin},
			\citenamefont {Bartenstein}, \citenamefont {Altmeyer}, \citenamefont {Riedl},
			\citenamefont {Jochim}, \citenamefont {Denschlag},\ and\ \citenamefont
			{Grimm}}]{Chin1128}%
		\BibitemOpen
		\bibfield  {author} {\bibinfo {author} {\bibfnamefont {C.}~\bibnamefont
				{Chin}}, \bibinfo {author} {\bibfnamefont {M.}~\bibnamefont {Bartenstein}},
			\bibinfo {author} {\bibfnamefont {A.}~\bibnamefont {Altmeyer}}, \bibinfo
			{author} {\bibfnamefont {S.}~\bibnamefont {Riedl}}, \bibinfo {author}
			{\bibfnamefont {S.}~\bibnamefont {Jochim}}, \bibinfo {author} {\bibfnamefont
				{J.~H.}\ \bibnamefont {Denschlag}},\ and\ \bibinfo {author} {\bibfnamefont
				{R.}~\bibnamefont {Grimm}},\ }\bibfield  {title} {\bibinfo {title}
			{\textrm{Observation of the Pairing Gap in a Strongly Interacting Fermi
					Gas}},\ }\href {https://doi.org/10.1126/science.1100818} {\bibfield
			{journal} {\bibinfo  {journal} {Science}\ }\textbf {\bibinfo {volume}
				{305}},\ \bibinfo {pages} {1128} (\bibinfo {year} {2004})}\BibitemShut
		{NoStop}%
		\bibitem [{\citenamefont {Regal}\ \emph {et~al.}(2004)\citenamefont {Regal},
			\citenamefont {Greiner},\ and\ \citenamefont {Jin}}]{PhysRevLett.92.040403}%
		\BibitemOpen
		\bibfield  {author} {\bibinfo {author} {\bibfnamefont {C.~A.}\ \bibnamefont
				{Regal}}, \bibinfo {author} {\bibfnamefont {M.}~\bibnamefont {Greiner}},\
			and\ \bibinfo {author} {\bibfnamefont {D.~S.}\ \bibnamefont {Jin}},\
		}\bibfield  {title} {\bibinfo {title} {\textrm{Observation of Resonance
					Condensation of Fermionic Atom Pairs}},\ }\href
		{https://doi.org/10.1103/PhysRevLett.92.040403} {\bibfield  {journal}
			{\bibinfo  {journal} {Phys. Rev. Lett.}\ }\textbf {\bibinfo {volume} {92}},\
			\bibinfo {pages} {040403} (\bibinfo {year} {2004})}\BibitemShut {NoStop}%
		\bibitem [{\citenamefont {Kraemer}\ \emph {et~al.}(2006)\citenamefont
			{Kraemer}, \citenamefont {Mark}, \citenamefont {Waldburger}, \citenamefont
			{Danzl}, \citenamefont {Chin}, \citenamefont {Engeser}, \citenamefont
			{Lange}, \citenamefont {Pilch}, \citenamefont {Jaakkola}, \citenamefont
			{N{\"a}gerl},\ and\ \citenamefont {Grimm}}]{CsEfi}%
		\BibitemOpen
		\bibfield  {author} {\bibinfo {author} {\bibfnamefont {T.}~\bibnamefont
				{Kraemer}}, \bibinfo {author} {\bibfnamefont {M.}~\bibnamefont {Mark}},
			\bibinfo {author} {\bibfnamefont {P.}~\bibnamefont {Waldburger}}, \bibinfo
			{author} {\bibfnamefont {J.~G.}\ \bibnamefont {Danzl}}, \bibinfo {author}
			{\bibfnamefont {C.}~\bibnamefont {Chin}}, \bibinfo {author} {\bibfnamefont
				{B.}~\bibnamefont {Engeser}}, \bibinfo {author} {\bibfnamefont {A.~D.}\
				\bibnamefont {Lange}}, \bibinfo {author} {\bibfnamefont {K.}~\bibnamefont
				{Pilch}}, \bibinfo {author} {\bibfnamefont {A.}~\bibnamefont {Jaakkola}},
			\bibinfo {author} {\bibfnamefont {H.-C.}\ \bibnamefont {N{\"a}gerl}},\ and\
			\bibinfo {author} {\bibfnamefont {R.}~\bibnamefont {Grimm}},\ }\bibfield
		{title} {\bibinfo {title} {\textrm{Evidence for Efimov quantum states in an
					ultracold gas of caesium atoms}},\ }\href@noop {} {\bibfield  {journal}
			{\bibinfo  {journal} {Nature 440, 315-318}\ } (\bibinfo {year}
			{2006})}\BibitemShut {NoStop}%
		\bibitem [{\citenamefont {Rui}\ \emph {et~al.}(2017)\citenamefont {Rui},
			\citenamefont {Yang}, \citenamefont {Liu}, \citenamefont {Zhang},
			\citenamefont {Liu}, \citenamefont {Nan}, \citenamefont {Chen}, \citenamefont
			{Zhao},\ and\ \citenamefont {Pan}}]{Shanghai2}%
		\BibitemOpen
		\bibfield  {author} {\bibinfo {author} {\bibfnamefont {J.}~\bibnamefont
				{Rui}}, \bibinfo {author} {\bibfnamefont {H.}~\bibnamefont {Yang}}, \bibinfo
			{author} {\bibfnamefont {L.}~\bibnamefont {Liu}}, \bibinfo {author}
			{\bibfnamefont {D.-C.}\ \bibnamefont {Zhang}}, \bibinfo {author}
			{\bibfnamefont {Y.-X.}\ \bibnamefont {Liu}}, \bibinfo {author} {\bibfnamefont
				{J.}~\bibnamefont {Nan}}, \bibinfo {author} {\bibfnamefont {Y.-A.}\
				\bibnamefont {Chen}}, \bibinfo {author} {\bibfnamefont {B.}~\bibnamefont
				{Zhao}},\ and\ \bibinfo {author} {\bibfnamefont {J.-W.}\ \bibnamefont
				{Pan}},\ }\bibfield  {title} {\bibinfo {title} {\textrm{Controlled
					state-to-state atom-exchange reaction in an ultracold atom-dimer mixture}},\
		}\href {http://dx.doi.org/10.1038/nphys4095} {\ \textbf {\bibinfo {volume}
				{13}},\ \bibinfo {pages} {699} (\bibinfo {year} {2017})}\BibitemShut
		{NoStop}%
		\bibitem [{\citenamefont {Ni}\ \emph {et~al.}(2008)\citenamefont {Ni},
			\citenamefont {Ospelkaus}, \citenamefont {de~Miranda}, \citenamefont {Peer},
			\citenamefont {Neyenhuis}, \citenamefont {Zirbel}, \citenamefont
			{Kotochigova}, \citenamefont {Julienne}, \citenamefont {Jin},\ and\
			\citenamefont {Ye}}]{KRb1}%
		\BibitemOpen
		\bibfield  {author} {\bibinfo {author} {\bibfnamefont {K.-K.}\ \bibnamefont
				{Ni}}, \bibinfo {author} {\bibfnamefont {S.}~\bibnamefont {Ospelkaus}},
			\bibinfo {author} {\bibfnamefont {M.~H.~G.}\ \bibnamefont {de~Miranda}},
			\bibinfo {author} {\bibfnamefont {A.}~\bibnamefont {Peer}}, \bibinfo {author}
			{\bibfnamefont {B.}~\bibnamefont {Neyenhuis}}, \bibinfo {author}
			{\bibfnamefont {J.~J.}\ \bibnamefont {Zirbel}}, \bibinfo {author}
			{\bibfnamefont {S.}~\bibnamefont {Kotochigova}}, \bibinfo {author}
			{\bibfnamefont {P.~S.}\ \bibnamefont {Julienne}}, \bibinfo {author}
			{\bibfnamefont {D.~S.}\ \bibnamefont {Jin}},\ and\ \bibinfo {author}
			{\bibfnamefont {J.}~\bibnamefont {Ye}},\ }\bibfield  {title} {\bibinfo
			{title} {\textrm{A High Phase-Space-Density Gas of Polar Molecules}},\ }\href
		{https://doi.org/10.1126/science.1163861} {\bibfield  {journal} {\bibinfo
				{journal} {Science}\ }\textbf {\bibinfo {volume} {322}},\ \bibinfo {pages}
			{231} (\bibinfo {year} {2008})}\BibitemShut {NoStop}%
		\bibitem [{\citenamefont {Ospelkaus}\ \emph {et~al.}(2010)\citenamefont
			{Ospelkaus}, \citenamefont {Ni}, \citenamefont {Wang}, \citenamefont
			{de~Miranda}, \citenamefont {Neyenhuis}, \citenamefont {Quemener},
			\citenamefont {Julienne}, \citenamefont {Bohn}, \citenamefont {Jin},\ and\
			\citenamefont {Ye}}]{Ospelkaus2010}%
		\BibitemOpen
		\bibfield  {author} {\bibinfo {author} {\bibfnamefont {S.}~\bibnamefont
				{Ospelkaus}}, \bibinfo {author} {\bibfnamefont {K.-K.}\ \bibnamefont {Ni}},
			\bibinfo {author} {\bibfnamefont {D.}~\bibnamefont {Wang}}, \bibinfo {author}
			{\bibfnamefont {M.~H.~G.}\ \bibnamefont {de~Miranda}}, \bibinfo {author}
			{\bibfnamefont {B.}~\bibnamefont {Neyenhuis}}, \bibinfo {author}
			{\bibfnamefont {G.}~\bibnamefont {Quemener}}, \bibinfo {author}
			{\bibfnamefont {P.~S.}\ \bibnamefont {Julienne}}, \bibinfo {author}
			{\bibfnamefont {J.~L.}\ \bibnamefont {Bohn}}, \bibinfo {author}
			{\bibfnamefont {D.~S.}\ \bibnamefont {Jin}},\ and\ \bibinfo {author}
			{\bibfnamefont {J.}~\bibnamefont {Ye}},\ }\bibfield  {title} {\bibinfo
			{title} {\textrm{Quantum-State Controlled Chemical Reactions of Ultracold
					Potassium-Rubidium Molecules}},\ }\href
		{https://doi.org/10.1126/science.1184121} {\bibfield  {journal} {\bibinfo
				{journal} {Science}\ }\textbf {\bibinfo {volume} {327}},\ \bibinfo {pages}
			{853} (\bibinfo {year} {2010})}\BibitemShut {NoStop}%
		\bibitem [{\citenamefont {Yan}\ \emph {et~al.}(2013)\citenamefont {Yan},
			\citenamefont {Moses}, \citenamefont {Gadway}, \citenamefont {Covey},
			\citenamefont {Hazzard}, \citenamefont {Rey}, \citenamefont {Jin},\ and\
			\citenamefont {Ye}}]{KRbSpin}%
		\BibitemOpen
		\bibfield  {author} {\bibinfo {author} {\bibfnamefont {B.}~\bibnamefont
				{Yan}}, \bibinfo {author} {\bibfnamefont {S.}~\bibnamefont {Moses}}, \bibinfo
			{author} {\bibfnamefont {B.}~\bibnamefont {Gadway}}, \bibinfo {author}
			{\bibfnamefont {J.}~\bibnamefont {Covey}}, \bibinfo {author} {\bibfnamefont
				{K.}~\bibnamefont {Hazzard}}, \bibinfo {author} {\bibfnamefont
				{A.}~\bibnamefont {Rey}}, \bibinfo {author} {\bibfnamefont {D.}~\bibnamefont
				{Jin}},\ and\ \bibinfo {author} {\bibfnamefont {J.}~\bibnamefont {Ye}},\
		}\bibfield  {title} {\bibinfo {title} {\textrm{Observation of dipolar
					spin-exchange interactions with lattice-confined polar molecules}},\
		}\href@noop {} {\bibfield  {journal} {\bibinfo  {journal} {Nature 501,
					521-525}\ } (\bibinfo {year} {2013})}\BibitemShut {NoStop}%
		\bibitem [{\citenamefont {DeMille}(2002)}]{DeMilleQuantComp}%
		\BibitemOpen
		\bibfield  {author} {\bibinfo {author} {\bibfnamefont {D.}~\bibnamefont
				{DeMille}},\ }\bibfield  {title} {\bibinfo {title} {\textrm{Quantum
					Computation with Trapped Polar Molecules}},\ }\href
		{https://doi.org/10.1103/PhysRevLett.88.067901} {\bibfield  {journal}
			{\bibinfo  {journal} {Phys. Rev. Lett.}\ }\textbf {\bibinfo {volume} {88}},\
			\bibinfo {pages} {067901} (\bibinfo {year} {2002})}\BibitemShut {NoStop}%
		\bibitem [{\citenamefont {Ospelkaus}\ \emph {et~al.}(2006)\citenamefont
			{Ospelkaus}, \citenamefont {Ospelkaus}, \citenamefont {Humbert},
			\citenamefont {Ernst}, \citenamefont {Sengstock},\ and\ \citenamefont
			{Bongs}}]{PhysRevLett.97.120402}%
		\BibitemOpen
		\bibfield  {author} {\bibinfo {author} {\bibfnamefont {C.}~\bibnamefont
				{Ospelkaus}}, \bibinfo {author} {\bibfnamefont {S.}~\bibnamefont
				{Ospelkaus}}, \bibinfo {author} {\bibfnamefont {L.}~\bibnamefont {Humbert}},
			\bibinfo {author} {\bibfnamefont {P.}~\bibnamefont {Ernst}}, \bibinfo
			{author} {\bibfnamefont {K.}~\bibnamefont {Sengstock}},\ and\ \bibinfo
			{author} {\bibfnamefont {K.}~\bibnamefont {Bongs}},\ }\bibfield  {title}
		{\bibinfo {title} {\textrm{Ultracold Heteronuclear Molecules in a 3D Optical
					Lattice}},\ }\href {https://doi.org/10.1103/PhysRevLett.97.120402} {\bibfield
			{journal} {\bibinfo  {journal} {Phys. Rev. Lett.}\ }\textbf {\bibinfo
				{volume} {97}},\ \bibinfo {pages} {120402} (\bibinfo {year}
			{2006})}\BibitemShut {NoStop}%
		\bibitem [{\citenamefont {Heo}\ \emph {et~al.}(2012)\citenamefont {Heo},
			\citenamefont {Wang}, \citenamefont {Christensen}, \citenamefont {Rvachov},
			\citenamefont {Cotta}, \citenamefont {Choi}, \citenamefont {Lee},\ and\
			\citenamefont {Ketterle}}]{GsDiMo23Na6LiBos2012}%
		\BibitemOpen
		\bibfield  {author} {\bibinfo {author} {\bibfnamefont {M.-S.}\ \bibnamefont
				{Heo}}, \bibinfo {author} {\bibfnamefont {T.~T.}\ \bibnamefont {Wang}},
			\bibinfo {author} {\bibfnamefont {C.~A.}\ \bibnamefont {Christensen}},
			\bibinfo {author} {\bibfnamefont {T.~M.}\ \bibnamefont {Rvachov}}, \bibinfo
			{author} {\bibfnamefont {D.~A.}\ \bibnamefont {Cotta}}, \bibinfo {author}
			{\bibfnamefont {J.-H.}\ \bibnamefont {Choi}}, \bibinfo {author}
			{\bibfnamefont {Y.-R.}\ \bibnamefont {Lee}},\ and\ \bibinfo {author}
			{\bibfnamefont {W.}~\bibnamefont {Ketterle}},\ }\bibfield  {title} {\bibinfo
			{title} {\textrm{Formation of ultracold fermionic NaLi Feshbach molecules}},\
		}\href {https://doi.org/10.1103/PhysRevA.86.021602} {\bibfield  {journal}
			{\bibinfo  {journal} {Phys. Rev. A}\ }\textbf {\bibinfo {volume} {86}},\
			\bibinfo {pages} {021602} (\bibinfo {year} {2012})}\BibitemShut {NoStop}%
		\bibitem [{\citenamefont {Rvachov}\ \emph {et~al.}(2017)\citenamefont
			{Rvachov}, \citenamefont {Son}, \citenamefont {Sommer}, \citenamefont
			{Ebadi}, \citenamefont {Park}, \citenamefont {Zwierlein}, \citenamefont
			{Ketterle},\ and\ \citenamefont {Jamison}}]{GsDiMo23Na6Li2017}%
		\BibitemOpen
		\bibfield  {author} {\bibinfo {author} {\bibfnamefont {T.~M.}\ \bibnamefont
				{Rvachov}}, \bibinfo {author} {\bibfnamefont {H.}~\bibnamefont {Son}},
			\bibinfo {author} {\bibfnamefont {A.~T.}\ \bibnamefont {Sommer}}, \bibinfo
			{author} {\bibfnamefont {S.}~\bibnamefont {Ebadi}}, \bibinfo {author}
			{\bibfnamefont {J.~J.}\ \bibnamefont {Park}}, \bibinfo {author}
			{\bibfnamefont {M.~W.}\ \bibnamefont {Zwierlein}}, \bibinfo {author}
			{\bibfnamefont {W.}~\bibnamefont {Ketterle}},\ and\ \bibinfo {author}
			{\bibfnamefont {A.~O.}\ \bibnamefont {Jamison}},\ }\bibfield  {title}
		{\bibinfo {title} {\textrm{Long-Lived Ultracold Molecules with Electric and
					Magnetic Dipole Moments}},\ }\href
		{https://doi.org/10.1103/PhysRevLett.119.143001} {\bibfield  {journal}
			{\bibinfo  {journal} {Phys. Rev. Lett.}\ }\textbf {\bibinfo {volume} {119}},\
			\bibinfo {pages} {143001} (\bibinfo {year} {2017})}\BibitemShut {NoStop}%
		\bibitem [{\citenamefont {Wu}\ \emph {et~al.}(2012)\citenamefont {Wu},
			\citenamefont {Park}, \citenamefont {Ahmadi}, \citenamefont {Will},\ and\
			\citenamefont {Zwierlein}}]{FBMol23Na40K2012}%
		\BibitemOpen
		\bibfield  {author} {\bibinfo {author} {\bibfnamefont {C.-H.}\ \bibnamefont
				{Wu}}, \bibinfo {author} {\bibfnamefont {J.~W.}\ \bibnamefont {Park}},
			\bibinfo {author} {\bibfnamefont {P.}~\bibnamefont {Ahmadi}}, \bibinfo
			{author} {\bibfnamefont {S.}~\bibnamefont {Will}},\ and\ \bibinfo {author}
			{\bibfnamefont {M.~W.}\ \bibnamefont {Zwierlein}},\ }\bibfield  {title}
		{\bibinfo {title} {\textrm{Ultracold Fermionic Feshbach Molecules of
					$^{23}\mathrm{Na}^{40}\mathrm{K}$}},\ }\href@noop {} {\bibfield  {journal}
			{\bibinfo  {journal} {Phys. Rev. Lett.}\ }\textbf {\bibinfo {volume} {109}},\
			\bibinfo {pages} {085301} (\bibinfo {year} {2012})}\BibitemShut {NoStop}%
		\bibitem [{\citenamefont {Park}\ \emph {et~al.}(2015)\citenamefont {Park},
			\citenamefont {Will},\ and\ \citenamefont {Zwierlein}}]{GsDiMo23Na40K2015}%
		\BibitemOpen
		\bibfield  {author} {\bibinfo {author} {\bibfnamefont {J.~W.}\ \bibnamefont
				{Park}}, \bibinfo {author} {\bibfnamefont {S.~A.}\ \bibnamefont {Will}},\
			and\ \bibinfo {author} {\bibfnamefont {M.~W.}\ \bibnamefont {Zwierlein}},\
		}\bibfield  {title} {\bibinfo {title} {\textrm{Ultracold Dipolar Gas of
					Fermionic $^{23}\mathrm{Na}^{40}\mathrm{K}$ Molecules in Their Absolute
					Ground State}},\ }\href {https://doi.org/10.1103/PhysRevLett.114.205302}
		{\bibfield  {journal} {\bibinfo  {journal} {Phys. Rev. Lett.}\ }\textbf
			{\bibinfo {volume} {114}},\ \bibinfo {pages} {205302} (\bibinfo {year}
			{2015})}\BibitemShut {NoStop}%
		\bibitem [{\citenamefont {Takekoshi}\ \emph {et~al.}(2012)\citenamefont
			{Takekoshi}, \citenamefont {Debatin}, \citenamefont {Rameshan}, \citenamefont
			{Ferlaino}, \citenamefont {Grimm}, \citenamefont {N\"agerl}, \citenamefont
			{Le~Sueur}, \citenamefont {Hutson}, \citenamefont {Julienne}, \citenamefont
			{Kotochigova},\ and\ \citenamefont {Tiemann}}]{PhysRevA.85.032506}%
		\BibitemOpen
		\bibfield  {author} {\bibinfo {author} {\bibfnamefont {T.}~\bibnamefont
				{Takekoshi}}, \bibinfo {author} {\bibfnamefont {M.}~\bibnamefont {Debatin}},
			\bibinfo {author} {\bibfnamefont {R.}~\bibnamefont {Rameshan}}, \bibinfo
			{author} {\bibfnamefont {F.}~\bibnamefont {Ferlaino}}, \bibinfo {author}
			{\bibfnamefont {R.}~\bibnamefont {Grimm}}, \bibinfo {author} {\bibfnamefont
				{H.-C.}\ \bibnamefont {N\"agerl}}, \bibinfo {author} {\bibfnamefont {C.~R.}\
				\bibnamefont {Le~Sueur}}, \bibinfo {author} {\bibfnamefont {J.~M.}\
				\bibnamefont {Hutson}}, \bibinfo {author} {\bibfnamefont {P.~S.}\
				\bibnamefont {Julienne}}, \bibinfo {author} {\bibfnamefont {S.}~\bibnamefont
				{Kotochigova}},\ and\ \bibinfo {author} {\bibfnamefont {E.}~\bibnamefont
				{Tiemann}},\ }\bibfield  {title} {\bibinfo {title} {\textrm{Towards the
					production of ultracold ground-state RbCs molecules: Feshbach resonances,
					weakly bound states, and the coupled-channel model}},\ }\href
		{https://doi.org/10.1103/PhysRevA.85.032506} {\bibfield  {journal} {\bibinfo
				{journal} {Phys. Rev. A}\ }\textbf {\bibinfo {volume} {85}},\ \bibinfo
			{pages} {032506} (\bibinfo {year} {2012})}\BibitemShut {NoStop}%
		\bibitem [{\citenamefont {Takekoshi}\ \emph {et~al.}(2014)\citenamefont
			{Takekoshi}, \citenamefont {Reichs\"ollner}, \citenamefont {Schindewolf},
			\citenamefont {Hutson}, \citenamefont {Le~Sueur}, \citenamefont {Dulieu},
			\citenamefont {Ferlaino}, \citenamefont {Grimm},\ and\ \citenamefont
			{N\"agerl}}]{GsDiMo87Rb133Cs2014Grimm}%
		\BibitemOpen
		\bibfield  {author} {\bibinfo {author} {\bibfnamefont {T.}~\bibnamefont
				{Takekoshi}}, \bibinfo {author} {\bibfnamefont {L.}~\bibnamefont
				{Reichs\"ollner}}, \bibinfo {author} {\bibfnamefont {A.}~\bibnamefont
				{Schindewolf}}, \bibinfo {author} {\bibfnamefont {J.~M.}\ \bibnamefont
				{Hutson}}, \bibinfo {author} {\bibfnamefont {C.~R.}\ \bibnamefont
				{Le~Sueur}}, \bibinfo {author} {\bibfnamefont {O.}~\bibnamefont {Dulieu}},
			\bibinfo {author} {\bibfnamefont {F.}~\bibnamefont {Ferlaino}}, \bibinfo
			{author} {\bibfnamefont {R.}~\bibnamefont {Grimm}},\ and\ \bibinfo {author}
			{\bibfnamefont {H.-C.}\ \bibnamefont {N\"agerl}},\ }\bibfield  {title}
		{\bibinfo {title} {\textrm{Ultracold Dense Samples of Dipolar \text{RbCs}
					Molecules in the Rovibrational and Hyperfine Ground State}},\ }\href
		{https://doi.org/10.1103/PhysRevLett.113.205301} {\bibfield  {journal}
			{\bibinfo  {journal} {Phys. Rev. Lett.}\ }\textbf {\bibinfo {volume} {113}},\
			\bibinfo {pages} {205301} (\bibinfo {year} {2014})}\BibitemShut {NoStop}%
		\bibitem [{\citenamefont {Wang}\ \emph {et~al.}(2015)\citenamefont {Wang},
			\citenamefont {He}, \citenamefont {Li}, \citenamefont {Zhu}, \citenamefont
			{Chen},\ and\ \citenamefont {Wang}}]{NaRbFeshbach}%
		\BibitemOpen
		\bibfield  {author} {\bibinfo {author} {\bibfnamefont {F.}~\bibnamefont
				{Wang}}, \bibinfo {author} {\bibfnamefont {X.}~\bibnamefont {He}}, \bibinfo
			{author} {\bibfnamefont {X.}~\bibnamefont {Li}}, \bibinfo {author}
			{\bibfnamefont {B.}~\bibnamefont {Zhu}}, \bibinfo {author} {\bibfnamefont
				{J.}~\bibnamefont {Chen}},\ and\ \bibinfo {author} {\bibfnamefont
				{D.}~\bibnamefont {Wang}},\ }\bibfield  {title} {\bibinfo {title}
			{\textrm{Formation of ultracold NaRb Feshbach molecules}},\ }\href
		{https://doi.org/10.1088/1367-2630/17/3/035003} {\bibfield  {journal}
			{\bibinfo  {journal} {New Journal of Physics}\ }\textbf {\bibinfo {volume}
				{17}},\ \bibinfo {pages} {035003} (\bibinfo {year} {2015})}\BibitemShut
		{NoStop}%
		\bibitem [{\citenamefont {Guo}\ \emph {et~al.}(2016)\citenamefont {Guo},
			\citenamefont {Zhu}, \citenamefont {Lu}, \citenamefont {Ye}, \citenamefont
			{Wang}, \citenamefont {Vexiau}, \citenamefont {Bouloufa-Maafa}, \citenamefont
			{Qu{\'{e}}m{\'{e}}ner}, \citenamefont {Dulieu},\ and\ \citenamefont
			{Wang}}]{NaRb3}%
		\BibitemOpen
		\bibfield  {author} {\bibinfo {author} {\bibfnamefont {M.}~\bibnamefont
				{Guo}}, \bibinfo {author} {\bibfnamefont {B.}~\bibnamefont {Zhu}}, \bibinfo
			{author} {\bibfnamefont {B.}~\bibnamefont {Lu}}, \bibinfo {author}
			{\bibfnamefont {X.}~\bibnamefont {Ye}}, \bibinfo {author} {\bibfnamefont
				{F.}~\bibnamefont {Wang}}, \bibinfo {author} {\bibfnamefont {R.}~\bibnamefont
				{Vexiau}}, \bibinfo {author} {\bibfnamefont {N.}~\bibnamefont
				{Bouloufa-Maafa}}, \bibinfo {author} {\bibfnamefont {G.}~\bibnamefont
				{Qu{\'{e}}m{\'{e}}ner}}, \bibinfo {author} {\bibfnamefont {O.}~\bibnamefont
				{Dulieu}},\ and\ \bibinfo {author} {\bibfnamefont {D.}~\bibnamefont {Wang}},\
		}\bibfield  {title} {\bibinfo {title} {\textrm{Creation of an Ultracold Gas
					of Ground-State Dipolar $^{23}\textrm{Na}^{87}\textrm{Rb}$ Molecules}},\
		}\bibfield  {journal} {\bibinfo  {journal} {Physical Review Letters}\
		}\textbf {\bibinfo {volume} {116}},\ \href
		{https://doi.org/10.1103/physrevlett.116.205303}
		{10.1103/physrevlett.116.205303} (\bibinfo {year} {2016})\BibitemShut
		{NoStop}%
		\bibitem [{\citenamefont {Zuchowski}\ and\ \citenamefont
			{Hutson}(2010)}]{Inelastic2010}%
		\BibitemOpen
		\bibfield  {author} {\bibinfo {author} {\bibfnamefont {P.~S.}\ \bibnamefont
				{Zuchowski}}\ and\ \bibinfo {author} {\bibfnamefont {J.~M.}\ \bibnamefont
				{Hutson}},\ }\bibfield  {title} {\bibinfo {title} {\textrm{Reactions of
					ultracold alkali-metal dimers}},\ }\href@noop {} {\bibfield  {journal}
			{\bibinfo  {journal} {Phys. Rev. A}\ }\textbf {\bibinfo {volume} {81}},\
			\bibinfo {pages} {060703(R)} (\bibinfo {year} {2010})}\BibitemShut {NoStop}%
		\bibitem [{\citenamefont {Hartmann}\ \emph {et~al.}(2019)\citenamefont
			{Hartmann}, \citenamefont {Schulze}, \citenamefont {Voges}, \citenamefont
			{Gersema}, \citenamefont {Gempel}, \citenamefont {Tiemann}, \citenamefont
			{Zenesini},\ and\ \citenamefont {Ospelkaus}}]{Hartmann2019}%
		\BibitemOpen
		\bibfield  {author} {\bibinfo {author} {\bibfnamefont {T.}~\bibnamefont
				{Hartmann}}, \bibinfo {author} {\bibfnamefont {T.~A.}\ \bibnamefont
				{Schulze}}, \bibinfo {author} {\bibfnamefont {K.~K.}\ \bibnamefont {Voges}},
			\bibinfo {author} {\bibfnamefont {P.}~\bibnamefont {Gersema}}, \bibinfo
			{author} {\bibfnamefont {M.~W.}\ \bibnamefont {Gempel}}, \bibinfo {author}
			{\bibfnamefont {E.}~\bibnamefont {Tiemann}}, \bibinfo {author} {\bibfnamefont
				{A.}~\bibnamefont {Zenesini}},\ and\ \bibinfo {author} {\bibfnamefont
				{S.}~\bibnamefont {Ospelkaus}},\ }\bibfield  {title} {\bibinfo {title}
			{\textrm{Feshbach resonances in $^{23}\mathrm{Na}+^{39}\mathrm{K}$ mixtures
					and refined molecular potentials for the \text{NaK} molecule}},\ }\href
		{https://doi.org/10.1103/PhysRevA.99.032711} {\bibfield  {journal} {\bibinfo
				{journal} {Phys. Rev. A}\ }\textbf {\bibinfo {volume} {99}},\ \bibinfo
			{pages} {032711} (\bibinfo {year} {2019})}\BibitemShut {NoStop}%
		\bibitem [{\citenamefont {Schulze}\ \emph {et~al.}(2018)\citenamefont
			{Schulze}, \citenamefont {Hartmann}, \citenamefont {Voges}, \citenamefont
			{Gempel}, \citenamefont {Tiemann}, \citenamefont {Zenesini},\ and\
			\citenamefont {Ospelkaus}}]{SchulzeBEC2018}%
		\BibitemOpen
		\bibfield  {author} {\bibinfo {author} {\bibfnamefont {T.~A.}\ \bibnamefont
				{Schulze}}, \bibinfo {author} {\bibfnamefont {T.}~\bibnamefont {Hartmann}},
			\bibinfo {author} {\bibfnamefont {K.~K.}\ \bibnamefont {Voges}}, \bibinfo
			{author} {\bibfnamefont {M.~W.}\ \bibnamefont {Gempel}}, \bibinfo {author}
			{\bibfnamefont {E.}~\bibnamefont {Tiemann}}, \bibinfo {author} {\bibfnamefont
				{A.}~\bibnamefont {Zenesini}},\ and\ \bibinfo {author} {\bibfnamefont
				{S.}~\bibnamefont {Ospelkaus}},\ }\bibfield  {title} {\bibinfo {title}
			{\textrm{Feshbach spectroscopy and dual-species Bose-Einstein condensation of
					$^{23}\mathrm{Na}\text{\ensuremath{-}}^{39}\mathrm{K}$ mixtures}},\ }\href
		{https://doi.org/10.1103/PhysRevA.97.023623} {\bibfield  {journal} {\bibinfo
				{journal} {Phys. Rev. A}\ }\textbf {\bibinfo {volume} {97}},\ \bibinfo
			{pages} {023623} (\bibinfo {year} {2018})}\BibitemShut {NoStop}%
		\bibitem [{\citenamefont {Schulze}\ \emph {et~al.}(2013)\citenamefont
			{Schulze}, \citenamefont {Temelkov}, \citenamefont {Gempel}, \citenamefont
			{Hartmann}, \citenamefont {Kn\"ockel}, \citenamefont {Ospelkaus},\ and\
			\citenamefont {Tiemann}}]{Schulze2013}%
		\BibitemOpen
		\bibfield  {author} {\bibinfo {author} {\bibfnamefont {T.~A.}\ \bibnamefont
				{Schulze}}, \bibinfo {author} {\bibfnamefont {I.~I.}\ \bibnamefont
				{Temelkov}}, \bibinfo {author} {\bibfnamefont {M.~W.}\ \bibnamefont
				{Gempel}}, \bibinfo {author} {\bibfnamefont {T.}~\bibnamefont {Hartmann}},
			\bibinfo {author} {\bibfnamefont {H.}~\bibnamefont {Kn\"ockel}}, \bibinfo
			{author} {\bibfnamefont {S.}~\bibnamefont {Ospelkaus}},\ and\ \bibinfo
			{author} {\bibfnamefont {E.}~\bibnamefont {Tiemann}},\ }\bibfield  {title}
		{\bibinfo {title} {\textrm{Multichannel modeling and two-photon coherent
					transfer paths in NaK}},\ }\href {https://doi.org/10.1103/PhysRevA.88.023401}
		{\bibfield  {journal} {\bibinfo  {journal} {Phys. Rev. A}\ }\textbf {\bibinfo
				{volume} {88}},\ \bibinfo {pages} {023401} (\bibinfo {year}
			{2013})}\BibitemShut {NoStop}%
		\bibitem [{\citenamefont {Voges}\ \emph {et~al.}(2019)\citenamefont {Voges},
			\citenamefont {Gersema}, \citenamefont {Hartmann}, \citenamefont {Schulze},
			\citenamefont {Tiemann}, \citenamefont {Ospelkaus},\ and\ \citenamefont
			{Zenesini}}]{Voges2019}%
		\BibitemOpen
		\bibfield  {author} {\bibinfo {author} {\bibfnamefont {K.~K.}\ \bibnamefont
				{Voges}}, \bibinfo {author} {\bibfnamefont {P.}~\bibnamefont {Gersema}},
			\bibinfo {author} {\bibfnamefont {T.}~\bibnamefont {Hartmann}}, \bibinfo
			{author} {\bibfnamefont {T.~A.}\ \bibnamefont {Schulze}}, \bibinfo {author}
			{\bibfnamefont {E.}~\bibnamefont {Tiemann}}, \bibinfo {author} {\bibfnamefont
				{S.}~\bibnamefont {Ospelkaus}},\ and\ \bibinfo {author} {\bibfnamefont
				{A.}~\bibnamefont {Zenesini}},\ }\bibfield  {title} {\bibinfo {title}
			{\textrm{A pathway to ultracold bosonic $^{23}\textrm{Na}^{39}\textrm{K}$
					ground state molecules}},\ }\href {http://arxiv.org/abs/1910.13771}
		{\bibfield  {journal} {\bibinfo  {journal} {arXiv:1910.13771}\ } (\bibinfo
			{year} {2019})}\BibitemShut {NoStop}%
		\bibitem [{\citenamefont {Chin}\ \emph {et~al.}(2010)\citenamefont {Chin},
			\citenamefont {Grimm}, \citenamefont {Julienne},\ and\ \citenamefont
			{Tiesinga}}]{RevModPhysFesh}%
		\BibitemOpen
		\bibfield  {author} {\bibinfo {author} {\bibfnamefont {C.}~\bibnamefont
				{Chin}}, \bibinfo {author} {\bibfnamefont {R.}~\bibnamefont {Grimm}},
			\bibinfo {author} {\bibfnamefont {P.}~\bibnamefont {Julienne}},\ and\
			\bibinfo {author} {\bibfnamefont {E.}~\bibnamefont {Tiesinga}},\ }\bibfield
		{title} {\bibinfo {title} {\textrm{Feshbach resonances in ultracold gases}},\
		}\href {https://doi.org/10.1103/RevModPhys.82.1225} {\bibfield  {journal}
			{\bibinfo  {journal} {Rev. Mod. Phys.}\ }\textbf {\bibinfo {volume} {82}},\
			\bibinfo {pages} {1225} (\bibinfo {year} {2010})}\BibitemShut {NoStop}%
		\bibitem [{\citenamefont {Fu}\ \emph {et~al.}(2014)\citenamefont {Fu},
			\citenamefont {Huang}, \citenamefont {Meng}, \citenamefont {Wang},
			\citenamefont {Zhang}, \citenamefont {Zhang}, \citenamefont {Zhai},
			\citenamefont {Zhang},\ and\ \citenamefont {Zhang}}]{SpinOrbitFeshMole}%
		\BibitemOpen
		\bibfield  {author} {\bibinfo {author} {\bibfnamefont {Z.}~\bibnamefont
				{Fu}}, \bibinfo {author} {\bibfnamefont {L.}~\bibnamefont {Huang}}, \bibinfo
			{author} {\bibfnamefont {Z.}~\bibnamefont {Meng}}, \bibinfo {author}
			{\bibfnamefont {P.}~\bibnamefont {Wang}}, \bibinfo {author} {\bibfnamefont
				{L.}~\bibnamefont {Zhang}}, \bibinfo {author} {\bibfnamefont
				{S.}~\bibnamefont {Zhang}}, \bibinfo {author} {\bibfnamefont
				{H.}~\bibnamefont {Zhai}}, \bibinfo {author} {\bibfnamefont {P.}~\bibnamefont
				{Zhang}},\ and\ \bibinfo {author} {\bibfnamefont {J.}~\bibnamefont {Zhang}},\
		}\bibfield  {title} {\bibinfo {title} {\textrm{Production of Feshbach
					molecules induced by spin-orbit coupling in Fermi gases}},\ }\href@noop {}
		{\bibfield  {journal} {\bibinfo  {journal} {Nature Physics 10, 110-115}\ }
			(\bibinfo {year} {2014})}\BibitemShut {NoStop}%
		\bibitem [{\citenamefont {Schulze}(2018)}]{Schulze2018}%
		\BibitemOpen
		\bibfield  {author} {\bibinfo {author} {\bibfnamefont {T.~A.}\ \bibnamefont
				{Schulze}},\ }\href {https://doi.org/https://doi.org/10.15488/3429} {Ph.D.
			thesis},\ \bibinfo  {school} {Leibniz Universit\"at Hannover} (\bibinfo
		{year} {2018})\BibitemShut {NoStop}%
		\bibitem [{\citenamefont {Hartmann}(2018)}]{Hartmannphd}%
		\BibitemOpen
		\bibfield  {author} {\bibinfo {author} {\bibfnamefont {T.}~\bibnamefont
				{Hartmann}},\ }\href {https://doi.org/https://doi.org/10.15488/4699} {Ph.D.
			thesis},\ \bibinfo  {school} {Leibniz Universit\"at Hannover} (\bibinfo
		{year} {2018})\BibitemShut {NoStop}%
		\bibitem [{\citenamefont {Camparo}\ and\ \citenamefont
			{Frueholz}(1984)}]{ARPdressedAtomInterpret1984}%
		\BibitemOpen
		\bibfield  {author} {\bibinfo {author} {\bibfnamefont {J.~C.}\ \bibnamefont
				{Camparo}}\ and\ \bibinfo {author} {\bibfnamefont {R.~P.}\ \bibnamefont
				{Frueholz}},\ }\bibfield  {title} {\bibinfo {title} {\textrm{A dressed atom
					interpretation of adiabatic rapid passage}},\ }\href
		{http://stacks.iop.org/0022-3700/17/i=20/a=015} {\bibfield  {journal}
			{\bibinfo  {journal} {Journal of Physics B: Atomic and Molecular Physics}\
			}\textbf {\bibinfo {volume} {17}},\ \bibinfo {pages} {4169} (\bibinfo {year}
			{1984})}\BibitemShut {NoStop}%
		\bibitem [{pri()}]{privcomET}%
		\BibitemOpen
		\href@noop {} {}\bibinfo {note} {E. Tiemann, private
			communication}\BibitemShut {NoStop}%
		\bibitem [{GFe()}]{GFesh}%
		\BibitemOpen
		\href@noop {} {}\bibinfo {note} {For calculating the two-body overlap
			integral we use the equation $g_{i,j} = \frac{1}{N_iN_j} \int_{V} n_i n_j
			\mathrm{d}V$, where $i$ and $j$ denote the atom species or molecules and $V$
			the volume. We assume the densities $n$ to follow a Gaussian
			distribution.}\BibitemShut {Stop}%
		\bibitem [{\citenamefont {Vitanov}\ \emph {et~al.}(2017)\citenamefont
			{Vitanov}, \citenamefont {Rangelov}, \citenamefont {Shore},\ and\
			\citenamefont {Bergmann}}]{RevModPhys.89.015006}%
		\BibitemOpen
		\bibfield  {author} {\bibinfo {author} {\bibfnamefont {N.~V.}\ \bibnamefont
				{Vitanov}}, \bibinfo {author} {\bibfnamefont {A.~A.}\ \bibnamefont
				{Rangelov}}, \bibinfo {author} {\bibfnamefont {B.~W.}\ \bibnamefont
				{Shore}},\ and\ \bibinfo {author} {\bibfnamefont {K.}~\bibnamefont
				{Bergmann}},\ }\bibfield  {title} {\bibinfo {title} {\textrm{Stimulated Raman
					adiabatic passage in physics, chemistry, and beyond}},\ }\href
		{https://doi.org/10.1103/RevModPhys.89.015006} {\bibfield  {journal}
			{\bibinfo  {journal} {Rev. Mod. Phys.}\ }\textbf {\bibinfo {volume} {89}},\
			\bibinfo {pages} {015006} (\bibinfo {year} {2017})}\BibitemShut {NoStop}%
	\end{thebibliography}

	%

\end{document}